# Kinetic Turnover in the Early-Stage Nucleation of Multi-Shell Condensed Clusters

Kaicheng Zhu[1,2] and Haibin Su[1,2]*

[1] Department of Chemistry, The Hong Kong University of Science and Technology, Hong Kong
[2] IAS Center for AI for Scientific Discoveries, The Hong Kong University of Science and Technology, Hong Kong

* To whom correspondence should be addressed. Email: *haibinsu@ust.hk*

**Abstract:** Nucleation is a key rate-limiting process in phase transition and phase separation. Recent studies highlight a significant discrepancy between experimentally measured nucleation rates and theoretical predictions, particularly when dynamic structural reordering occurs along multi-step pathways. To bridge this gap, we develop a multi-shell model with a space-time-dependent order-parameter field to describe the reordering-nucleation process, where structural reorganization couples with the early growth of condensed clusters. Through stochastic simulations, we track the time-resolved evolution of heterogeneous structural order inside growing clusters. Path analysis of the first-passage problem in early-stage nucleation demonstrates that shifting the reordering rate alters the nucleation rate by several orders of magnitude. Furthermore, as the coupling strength increases, the relationship between the mean first-passage time and reordering susceptibility shifts from monotonic to non-monotonic, exhibiting a turnover effect. We quantitatively rationalize these behaviors with an effective nucleation barrier that accounts for non-equilibrium properties. Our findings elucidate the mechanisms behind multi-step nucleation and offer a predictive framework for future studies.

**Teaser:** Coupling structural reordering with surface growth induces a non-monotonic turnover in nucleation kinetics.

## INTRODUCTION

Understanding nucleation dynamics during non-equilibrium phase transition and phase separation is a crucial objective in physical science, particularly given the recent advances in nanocrystallization and biomolecular condensation(*1–6*). The nucleation process acts as the primary rate-limiting step in the early-stage development of finite-size phase domains, typically characterized by particle aggregation and the emergence of condensed clusters from a dilute background. The evolution of these condensed clusters traverses a series of unstable states before reaching a critical configuration where the net nucleation-driving force transitions from negative to positive. In the classical nucleation theory, the free energy difference between this critical state and the initial state is used to quantify the nucleation barrier and estimate the nucleation rate and flux(*7–9*). While non-classical nucleation theories have extended this paradigm by incorporating multi-step mechanisms(*10–18*), recent empirical studies have revealed substantial discrepancies between experimentally measured nucleation rates and theoretical predictions(*19–21*). Furthermore, although the coevolution of cluster size and structural motifs within the condensed phase has been widely observed, its precise impact on the dynamic properties of the nucleation process remains poorly understood(*22–29*).

High-resolution measurements have exposed three prominent features of structural order within condensed clusters. First, the internal structure of an individual cluster evolves dynamically across distinct states while its size changes over time. With single-particle resolution, researchers have observed structural transitions between states of different symmetries during liquid-solid phase separation in various colloidal systems(*26*). Second, structural motifs within a single cluster are inherently heterogeneous, exhibiting a non-uniform gradient from the cluster core to the interface. Using time-resolved atomic-resolution imaging, studies have shown that the structural-order gradient in nanocrystal is smooth near the center but steep near the interface, varying systematically with the cluster radius, a phenomenon also corroborated by recent molecular dynamics simulations(*27–30*). Third, the local structure at the interface directly dictates the cluster growth rate. Full-atom simulations have demonstrated a difference of over three orders of magnitude in the instantaneous growth rate depending on the interfacial structure, a consequence of altered activation barriers for surface adsorption(*31*). Although experimental limitations historically restricted these observations to the late, (meta-)stable growth stage, these structural features play a significant role during early-stage nucleation as well.

In this study, we develop a multi-shell model to describe the stochastic dynamics governing the coupled reordering-nucleation process (Fig. 1). The spatiotemporal structural order within a single condensed cluster is phenomenologically quantified by a non-conserved order-parameter field $\phi(\mathrm{x},\mathrm{t})$, constructed over a discretized multi-shell domain. The potential energy is formulated to account for both local intra-shell interaction and inter-shell coupling between adjacent shell layers. In contrast to existing theoretical frameworks(*32–36*), our model explicitly couples two distinct reaction mechanisms: intra-phase structural reordering and inter-phase surface adsorption. The underlying reaction probabilities are determined from transition-state barriers via Marcus theory(*37–39*), while transport timescales are absorbed into the waiting times between successive adsorption events. Through this mathematical formulation, we extend the conventional isolated critical point in the classical nucleation theory to a high-dimensional hypersurface, termed the critical nucleation regime. We investigate the global stochastic dynamics and the associated first-passage problem using numerical simulations and path analysis, utilizing the mean first-passage time to characterize the nucleation kinetics. Our findings reveal that the characteristic nucleation time varies by several orders of magnitude as the reordering susceptibility changes, a non-equilibrium phenomenon that escapes previous theoretical explanations. Additionally, under strong coupling conditions, the system exhibits a distinct turnover effect, where the relationship between the nucleation time and reordering susceptibility becomes fundamentally non-monotonic.

## MODEL

The concept of the reordering-nucleation process is extended from the classical nucleation theory and the two-step nucleation model in a multi-step fashion (Fig. 1a). The reordering-nucleation pathways span a large phase-space region of multiple degrees of freedom. The shifts along nucleation coordinates characterize the condensation steps driven by spatial transportation in the dilute background and adsorption reaction at the interface. The changes of structure coordinates reflect the fine-tuning of the spatial patterns of particles in the condensed phase, which are often measured with structural factors. Inspired by non-equilibrium statistical mechanics, we designed a field-theoretic formulation of the reordering-nucleation dynamics with two fields(*40–42*): the conserved density field $\rho(x,t)$ and the non-conserved structural order-parameter field $\phi(x,t)$. To perform an explicit analysis of the early nucleation dynamics, we invent the multi-shell model in this paper.

There are two key factors that motivate the construction of the multi-shell model for the reordering-nucleation process. First, the early growth of a cluster is mainly achieved through surface adsorption events, which dates back to Ostwald's ripening(*43*, *44*). At this stage, most of the particles are dissolved in the dilute background, and the number of condensed clusters is small. Thus, collisions between clusters are extremely rare compared to those between a cluster and dissolved particles. Based on this property, we neglect the cluster-cluster coalescence during early nucleation and absorb the diffusion-limited collision waiting time into the overall waiting time of adsorption $\tau_r$. Second, the interfacial interaction is strong when a cluster is small due to the large surface-to-volume ratio. This restricts the shape of a cluster to be approximately spherical. Moreover, the spatial pattern of structure motifs inside a single cluster also roughly preserves the rotational symmetry(*24*, *27*, *28*, *30*).

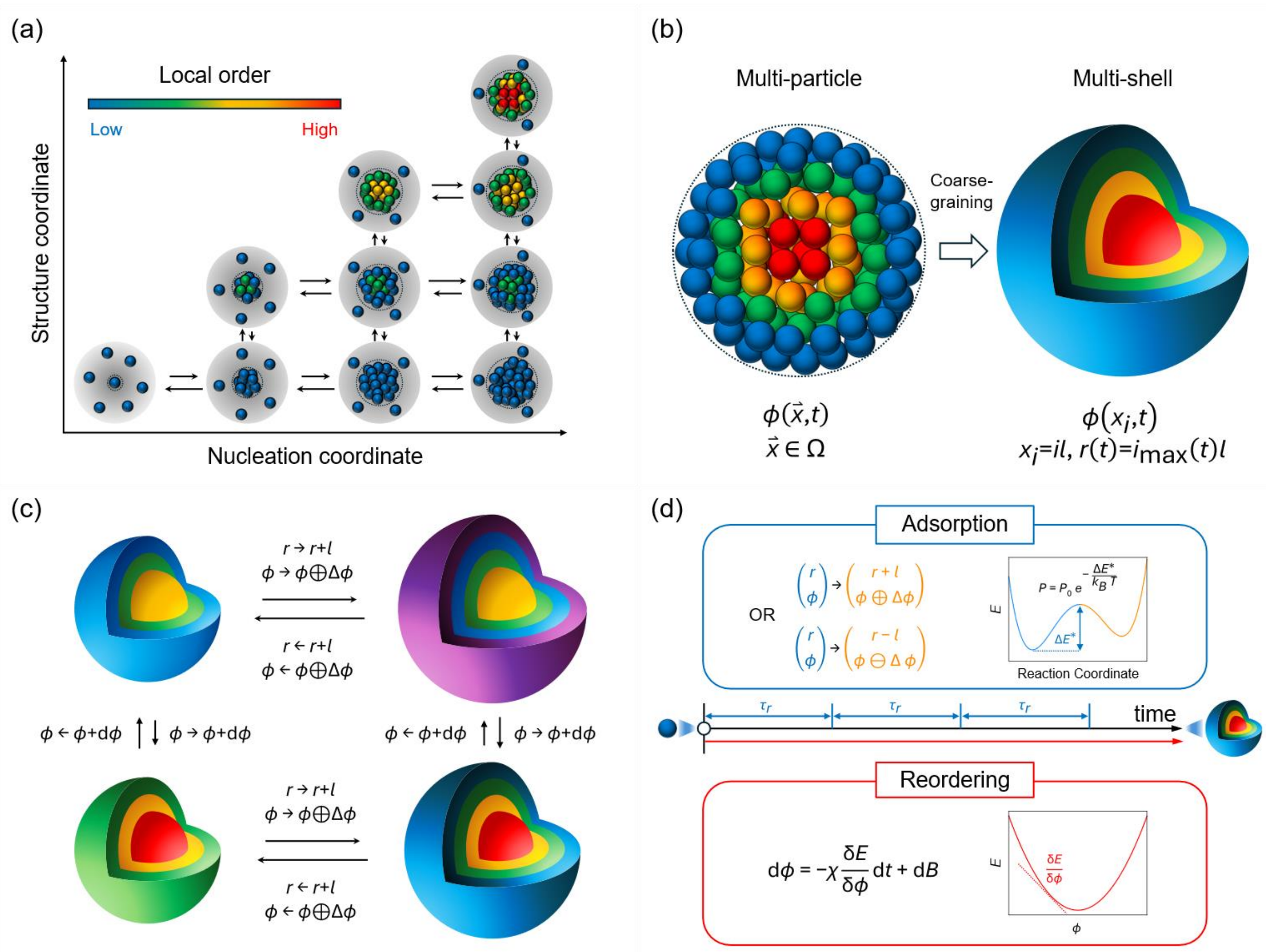

Figure 1. **Overview of the reordering-nucleation process within the multi-shell model. (a)** Schematic of the reordering-nucleation pathways projected onto the nucleation and structural coordinates. **(b)** Representation of the multi-shell model, providing a reduced spatial description of a single condensed cluster where the continuous space domain of the order-parameter field $\phi(x,t)$ is discretized by shell thickness $l$ and the cluster radius $r$. **(c)** Reaction mechanisms governing adsorption (horizontal) and reordering (vertical) events. **(d)** Multiscale dynamics coupling discrete adsorption and continuous reordering events. Adsorption steps occur discretely, separated by a characteristic waiting time $\tau_r$, with reaction probabilities determined by the transition-state barrier. In contrast, structural reordering evolves continuously over finite relaxation timescales.

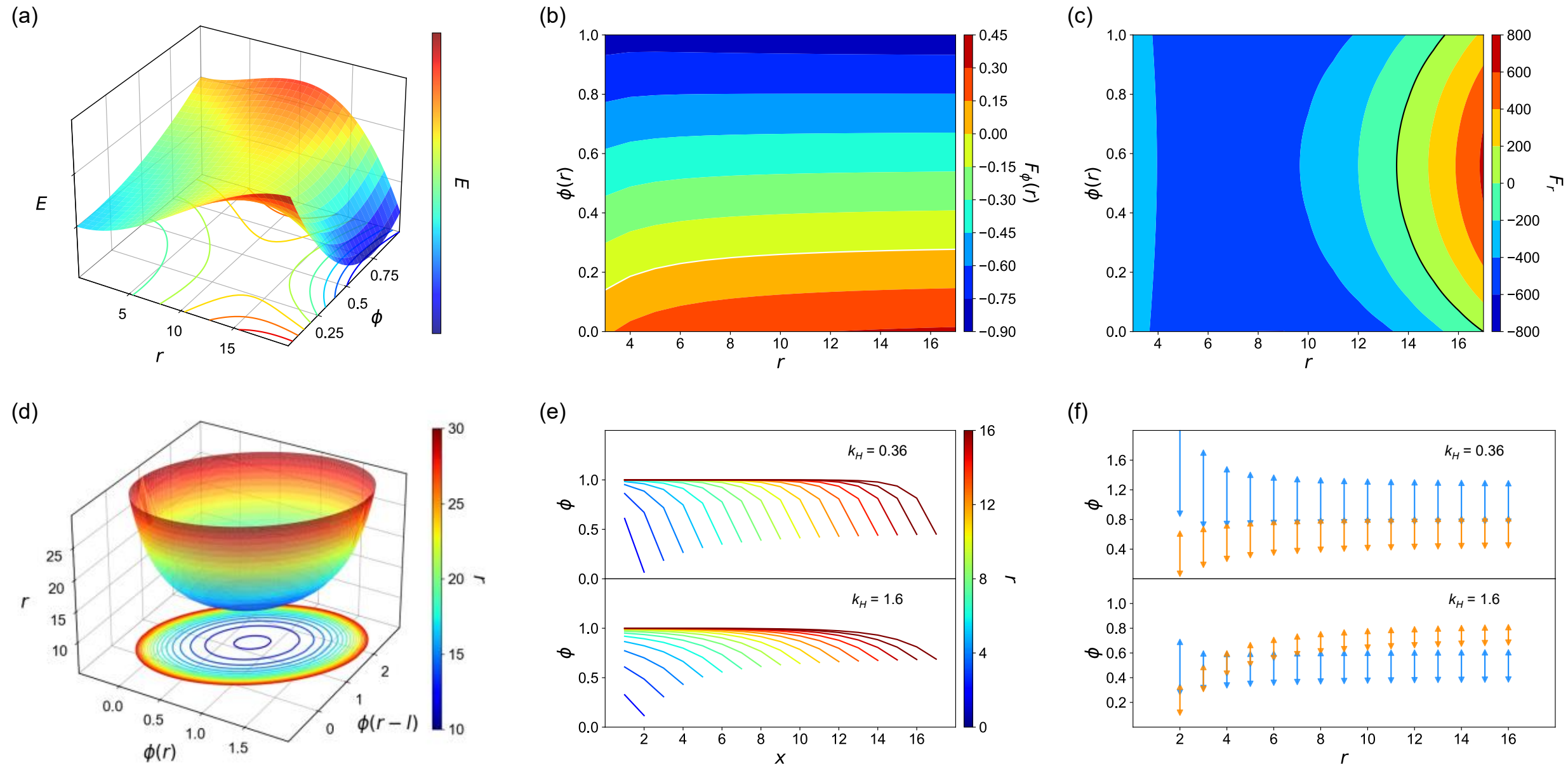

Figure 2. **Overview of the reaction mechanics. (a)** Schematic illustration of the $E$-energy landscape projected onto the $(r, \phi)$ phase space. **(b)** The $(r, \phi)$-dependent reordering-driving force $F_\phi$ acting at the interface when $\phi(r-l) = (\phi_S + \phi_C)/2$. **(c)** The $(r, \phi)$-dependent nucleation-driving force $F_r$ when $\phi(r-l) = (\phi_S + \phi_C)/2$. **(d)** The critical nucleation regime ($F_r = 0$) represented within the three-dimensional $[r, \phi(r), \phi(r-l)]$ phase space. $k_H = 0.36$ for panels (b), (c), and (d). **(e)** Size-dependent full-reordered states as a function of cluster radius. **(f)** Comparison between pro-nucleation states (blue) and full-reordered states (orange); the up triangles denote the corresponding $\phi(r-l)$ and the down triangles denote the corresponding $\phi(r)$.

We build a multi-shell model to reduce $\rho(x,t)$ to a scalar variable $r(t)$ and nest $\phi(x,t)$ in a discrete spatial domain (Fig. 1b). The order-parameter field is reformulated in a discrete fashion as $\phi(x_i, t)$, where $x_i$ is the radial coordinate of the shell indexed by $i \in \mathbb{Z}^+$. All shells have the same thickness $l$, so $x = il$ and $r = i_{max} l$. The spatial unit is normalized with respect to $l$, i.e., $l = 1$. In a short-hand form, $x = i$ and $r = i_{max}$. Because the radius changes with time, $r$ is a time-dependent variable that labels the outermost shell at the interface. With this construction, we define the potential energy of a single cluster as follows:

$$E(\phi) = \int v_S(\phi)\, dA + \int v_C(\phi)\, dV, \tag{1}$$

where the first integral on RHS is taken over the interfacial area and the second over the bulk domain inside the entire cluster. $v_{S,C}(\phi)$ are the corresponding potential energy densities, whose specific expressions are written as below:

$$v_S(\phi) = e_S + k_S|\phi - \phi_S|^2, \tag{2a}$$
$$v_C(\phi) = e_C + k_C|\phi - \phi_C|^2 + k_H|\nabla\phi|^2, \tag{2b}$$

where $e_{S,C}$ are bare local energy densities, $\phi_{S,C}$ are local order parameters for bare ground states ($\phi_S = 0$, and $\phi_C = 1$), $k_{S,C}$ are local coupling strengths, with the subscripts $S$ and $C$ denoting the interfacial and condensed regions, and $k_H$ is the shell-shell coupling strength. $\nabla\phi$ is the relative difference of adjacent-shell configurations (Methods). The coupling-induced energy shifts are schematically illustrated in Fig. 2a.

Two reaction mechanisms are considered: the intra-phase structural reordering and the inter-phase adsorption, combined into a composite stochastic process (Fig. 1c). In an elementary reordering event, the structural order in the condensed cluster changes while the radius is fixed. In a continuous fashion, this can be expressed as $\phi \to \phi + d\phi$. Since $\phi$ is a phenomenological characterization of the structural order at mesoscale, and there is no preserved structure in the dilute region, we set the background order parameter to the constant $\phi_S$, the same as the initial condition of a minimal cluster of $r = 1$, and do not treat the dilute-phase reordering explicitly. The mechanism of adsorption is quite different from reordering. In a forward adsorption step where $r \to r + l$, the order-parameter values over existing shells are fixed, but it is extended by an additional component $\phi(r + l) = \phi_S$, as the cluster gains a new outermost shell. In a backward adsorption step where $r \to r - l$, the order-parameter field is truncated at the interface and loses its component at $r$, while other components remain fixed. Thus, the set of shell indices is changed to $\{1,2, \dots, r - l\}$. The change of the order-parameter field during adsorption resembles the procedure of direct sum, so we use the symbolic notation $\phi \to \phi \oplus \Delta\phi$ for convenience. Because both $r$ and $\phi$ always change together during adsorption, they are not independent variables in classical statistical mechanics(*40*). We call this hidden connection between $r$ and $\phi$ as implicit coupling and express this relation as $\Delta\phi = \gamma\Delta r$, where $\gamma$ is the implicit coupling coefficient. Consequently, if no reordering occurs, $\phi(x) = \phi_S$ at all sites.

The reaction mechanisms involve distinct characteristic timescales (Fig. 1d). The reordering events describe local structural reorganization inside the condensed phase, and their waiting time $\tau_\phi$ is much shorter than adsorption waiting time $\tau_r$, due to the diffusion-limited transportation in the dilute phase. Therefore, $\tau_\phi \ll \tau_r$. Besides, the duration of an elementary reaction step is negligible compared with the waiting times. Since the overall nucleation kinetics are measured on the timescale much longer than that of a single adsorption event $\tau_r$, the change of $\phi$ induced by reordering can be formulated in a continuous manner as follows(*39*, *45–47*):

$$d\phi = \chi F_\phi dt + dB, \tag{3}$$

with the reordering-driving force

$$F_\phi = -\frac{\delta E}{\delta \phi}. \tag{4}$$

$dB$ denotes a Wiener process, $\chi$ is the reordering susceptibility as a rate parameter for the dynamic response of $\phi$ to $F_\phi$. The reordering-driving force is derived from a functional derivative via the variational principle(*48*). From Eqs. (1-2), $F_\phi$ is given as a linear form of $\phi$. Its $(r, \phi)$ dependence is captured by calculating the local values of $F_\phi$ at $x = r$ under different $r$ and $\phi(r)$ conditions while $\phi(r - l)$ is fixed as $(\phi_S + \phi_C)/2$ (Fig.2b).

Different from reordering, the adsorption events are sparsely distributed along the timeline, due to its longer waiting time $\tau_r$ compared to $\tau_\phi$ (Fig. 1d). In an elementary adsorption step, the reaction probability is restricted by the transition-state barrier $\Delta E^*$:

$$P_\pm = P_0 \exp\left(-\frac{\Delta E_\pm^*}{k_B T}\right), \tag{5}$$

where subscripts $\pm$ denote specific types of adsorption events as forward or backward. $P_0$ is a normalization factor, $k_B$ is the Boltzmann constant and $T$ is temperature. Since the microscopic mechanism of adsorption is non-adiabatic particle transfer between the dilute and condensed domains, we apply the Marcus theory for the transition-state barrier(*36–38*):

$$\Delta E_\pm^* = \frac{\lambda}{4}\left(1 + \frac{\Delta E_\pm}{\lambda}\right)^2, \tag{6}$$

where $\lambda$ is the reorganization energy, and $\Delta E_\pm$ is the energy gap between the product and reactant states in forward and backward adsorptions. We assume that $\lambda \gg \Delta E_\pm$ in all relevant adsorption events, so that the nucleation dynamics can be formulated as follows:

$$\Delta r = \mu e^{-\frac{\lambda}{4k_B T}} F_r \Delta t + \Delta_\xi r, \tag{7}$$

where $\mu$ is the mobility, $F_r$ is the nucleation-driving force, and $\Delta_\xi r$ denotes the random fluctuations in adsorption. Because the spatial domain is discretized with the shell thickness $l$, the nucleation-driving force is taken as follows:

$$F_r = -\frac{\Delta E_+ - \Delta E_-}{2l}. \tag{8}$$

In contrast to $F_\phi$ which is a space-dependent function that can be represented in a vector form, $F_r$ is a scalar. As adsorption steps are only relevant with the interface and near-interface regions, only local values of $\phi$ at the outermost and penultimate shells are involved in the nucleation-driving force. We examine the $(r, \phi)$ dependence of $F_r$ by fixing the value of $\phi(r - l)$ as $(\phi_S + \phi_C)/2$ while tuning $r$ and $\phi(r)$ (Fig. 2c). When $r$ is small, $F_r$ is mostly negative. There is a special regime where $F_r$ approaches zero, and if $r$ keeps increasing over this regime, $F_r$ turns positive. This picture indicates a generalization of the classical nucleation theory: this regime where $F_r = 0$ is an extension of the critical point in previous theories. Therefore, we call it the critical nucleation regime. Since there are only three relevant scalar components in the nucleation-driving force: $r$, $\phi(r)$, and $\phi(r - l)$, we project the critical nucleation regime onto the three-dimensional phase space of $[r, \phi(r), \phi(r - l)]$. This regime is a hypersurface in the multi-dimensional phase space (Fig. 2d). There exists an infimum radius below which $F_r$ is strictly negative, and this behavior correctly reflected the dynamic nature of early nucleation where the growth of condensed domains is driven by fluctuations rather than the net driving force.

We analyze two important types of states: full-reordered states $\phi_{FR}(x;r) \equiv \phi|_{F_{\phi=0}}$, and pro-nucleation states $\phi_{PN}(x;r) \equiv \phi|_{\delta F_r/\delta\phi=0}$. A full-reordered state is a state of vanishing reordering-driving force and is also the expected state when $\chi \to \infty$. The gradients of full-reordered states depend on the shell-shell coupling strength $k_H$, as it contributes to the heterogeneous part in the potential energy (Eq. 2b). The local order-parameter values of full-reordered states increase monotonically with radius (Fig. 2e), because surface-to-volume ratio decreases and $\phi_C$ is set larger than $\phi_S$. Two spatial features of $\phi_{FR}$ show good agreement with recent experiments(*27*, *28*): first, $\phi_{FR}$ generally approach $\phi_C$ (which is set to 1) as radius increases; second, the local gradients are smaller in the near-center region than in the near-interface region.

The pro-nucleation states are termed as they maximize the nucleation-driving force $F_r$. Because $F_r$ is determined by the outmost- and penultimate-shell states, the relevant components of $\phi_{PN}$ are those at the positions $x = r$ and $x = r - l$, i.e., only $\phi_{PN}(r)$ and $\phi_{PN}(x - l)$ are relevant (Fig. 2f). The relative differences between $\phi_{FR}$ and $\phi_{PN}$ depend on the shell-shell coupling strength $k_H$. In the weak-coupling case, $\phi_{PN} > \phi_{FR}$, but in the strong-coupling case, $\phi_{PN}$ are smaller than $\phi_{FR}$ after a threshold of radius. As $\phi(x) = \phi_S$ without reordering, $F_\phi$ continuously drives the $\phi$ field from $\phi_S$ (which is set to 0) to $\phi_{FR}$, so when $\phi_{PN}$ lies between $\phi_S$ and $\phi_{FR}$ at strong coupling, reordering would not monotonically favor the nucleation-driving force. This interesting property would further lead to a kinetic turnover of the first-passage time in early nucleation.

## EARLY NUCLEATION

The study of early nucleation focuses on the non-equilibrium evolution of a single cluster from the initial state to beyond the critical nucleation regime. Since this process covers the timescale much larger than the waiting times of reordering and adsorption events ($\tau_\phi$ and $\tau_r$), symbolically assembling the derivations and neglecting the intrinsic fluctuations in reordering give the stochastic equations of motion in a continuous-time fashion:

$$\frac{dr}{dt} = \mu e^{-\frac{\lambda}{4k_BT}} F_r + \xi_r, \tag{9a}$$

$$\frac{d\phi}{dt} = \chi F_\phi \oplus \gamma \frac{dr}{dt}, \tag{9b}$$

where $\xi_r$ is a Gaussian white noise and $\gamma$ is the implicit-coupling coefficient that addresses the co-change of $r$ and $\phi$ during adsorption. $\oplus$ indicates the appending or removing of components of the $\phi$ function at $x = r$.

We conducted the stochastic simulations of this reordering-nucleation dynamics under different parameter conditions (Methods). For every condition, total $N = 1000$ clusters were simulated. To ensure trajectories covered the complete early-stage nucleation, we examined, $N_{[F_r<0]}$, the number of sub-critical clusters whose nucleation-driving forces were negative. $N_{[F_r<0]}$ consistently decayed from $N$ to zero as time shifted from the beginning to the end.

With the simulation data, we performed the statistical analysis at separate times. The time-dependent radius distributions are shown in Fig. 3a. In early-stage nucleation, enlargement of the condensed clusters is characterized by the extension of distribution functions at the large-radius tails, while most clusters stay small. It implies that this process is unstable. As time goes on, rare and large clusters reach and pass the critical nucleation regime to enter the stable-growth regime where the nucleation-driving force is positive. Since $F_r$ is positive and it further increases with radius, the clusters would not return to the small-radius states once in the stable-growth regime. This leads to an irreversible transition from the early-stage nucleation to the late-stage growth. During this dynamic transition, the distribution functions dramatically change with the rising of a second peak at the large-radius regime, indicating the emergence of stable large clusters(*35*, *36*).

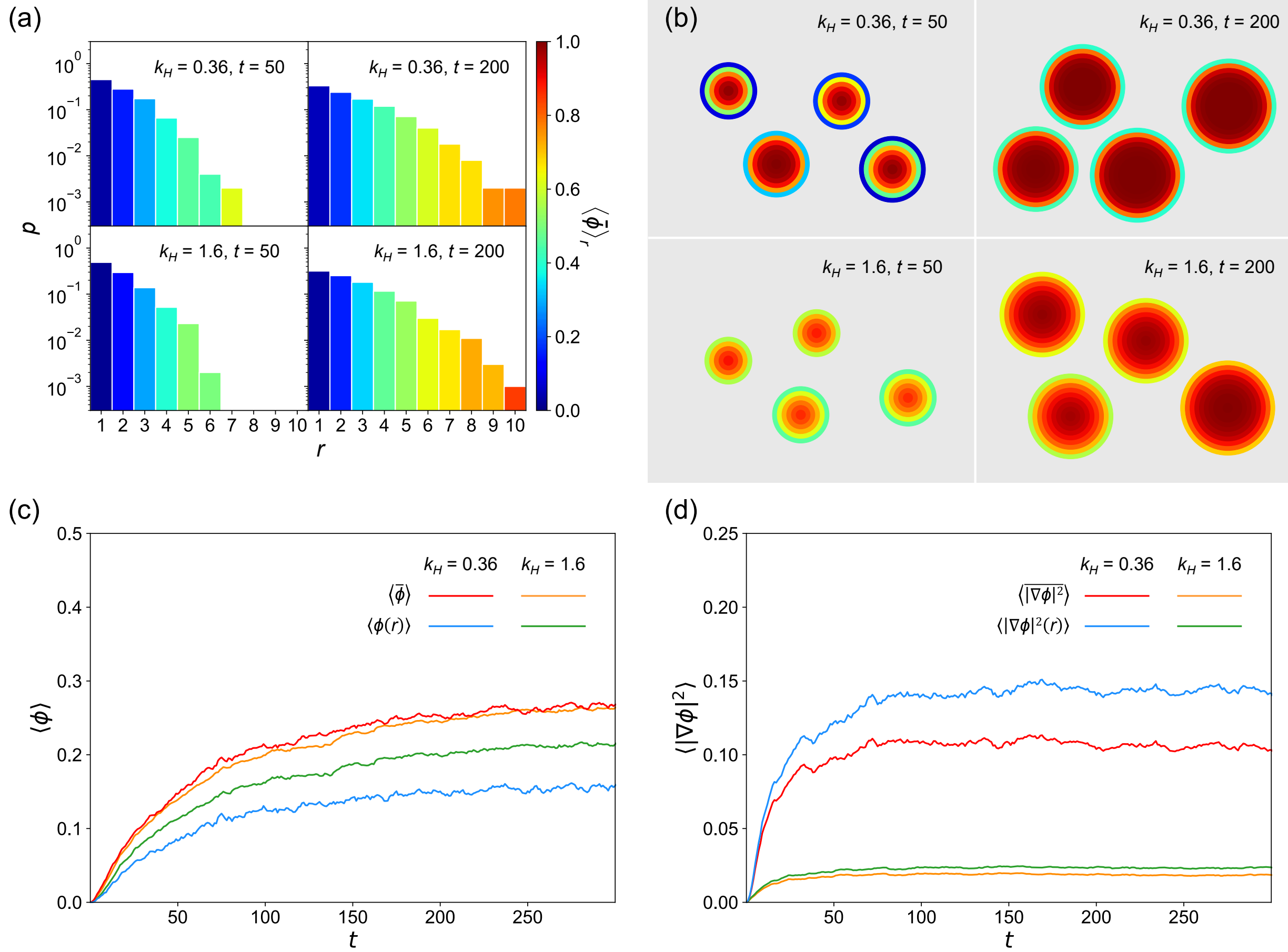

Figure 3. **Overview of early-stage nucleation dynamics. (a)** Cluster radius $r$ distributions at weak ($k_H$ = 0.36) and strong ($k_H$ = 1.6) couplings, captured at $t$ = 50 and 200, respectively. The color scheme indicates the spatially averaged order parameters $\langle\bar{\phi}\rangle_r$ corresponding to distinct radius groups. **(b)** Representative morphological configurations of the condensed clusters, projected onto 2D planes. **(c)** Time-resolved evolution of the ensemble-averaged order parameters (spatial average $\langle\bar{\phi}\rangle$ and interfacial value $\langle\phi(r)\rangle$). **(d)** Time-resolved evolution of the ensemble-averaged spatial gradients (spatial average $\langle\overline{|\nabla\phi|^2}\rangle$ and interfacial value $\langle|\nabla\phi|^2(r)\rangle$). Here, overlines and angle brackets denote spatial and ensemble averages, respectively. All data presented are obtained from simulations with the reordering susceptibility $\chi$ = 1.

The color scheme in Fig. 3a denotes the space-averaged order parameters associated with different radius groups: $\langle\bar{\phi}\rangle_r = \langle\frac{3}{r^3}\int \phi|_r x^2 dx\rangle_r$, where $\phi|_r$ denotes the order-parameter fields with the radius $r$, the bar notation stands for space average, and the labeled angle brackets indicate that the group average is evaluated among the related subset of the ensemble. This color scheme helps to illustrate the coupling between the radius and order parameters, consistent with the radius-dependent properties of the reordering-driving force. It is likely to see a larger cluster with a higher space-averaged order parameter than a smaller cluster. For clarity, we demonstrate the 2D projections of representative samples in Fig. 3b. The colored rings denote the shell layers with different local order parameters projected onto 2D space. The presented clusters are selected to stand for the largest ones in the corresponding ensembles at the same times as in Fig. 3a. The correlation between the cluster size and the internal structural order is vividly manifested: the large clusters tend to have higher order parameters, and vice versa. The non-uniform gradients of order-parameter fields extending from the centers to the interfaces are observed, and the gradients get steeper in the near-interface regions than the central regions. The amplitudes of the gradients generally decrease in large clusters. This feature is in good agreement with recent experiments and cannot be described by previous theories(*27*, *28*).

We computed the local values of $\phi$ and $|\nabla\phi|^2$ at the interface, with their spatial averages $\bar{\phi}$ and $\overline{|\nabla\phi|^2}$, and exhibited the time-resolved ensemble-averaged results (Fig. 3c and Fig. 3d). The ensemble averages are calculated from all clusters within the same simulation conditions, in contrast to the group averages calculated from radius-specific subsets (Fig. 3a). The evolution of order parameters is featured by the adaptation-like behaviors when the radii change, which is favored by the reordering-driving force. Compared with $\langle\bar{\phi}\rangle$, $\langle\phi(r)\rangle$ are closer to $\phi_S$, due to the local interaction at interface. This behavior is enhanced at weak coupling, because the interfacial state gets less affected by the internal states via shell-shell interaction. The interfacial gradients of order-parameter fields rise sharply as time increases, while the overall space-averaged gradients remain relatively small. In other words, the $\phi$ fields are more homogeneous inside and more heterogeneous outside, characterized by the increase of gradients from centers to interfaces. Also, the gradients tend to be higher with a smaller coupling strength $k_H$.

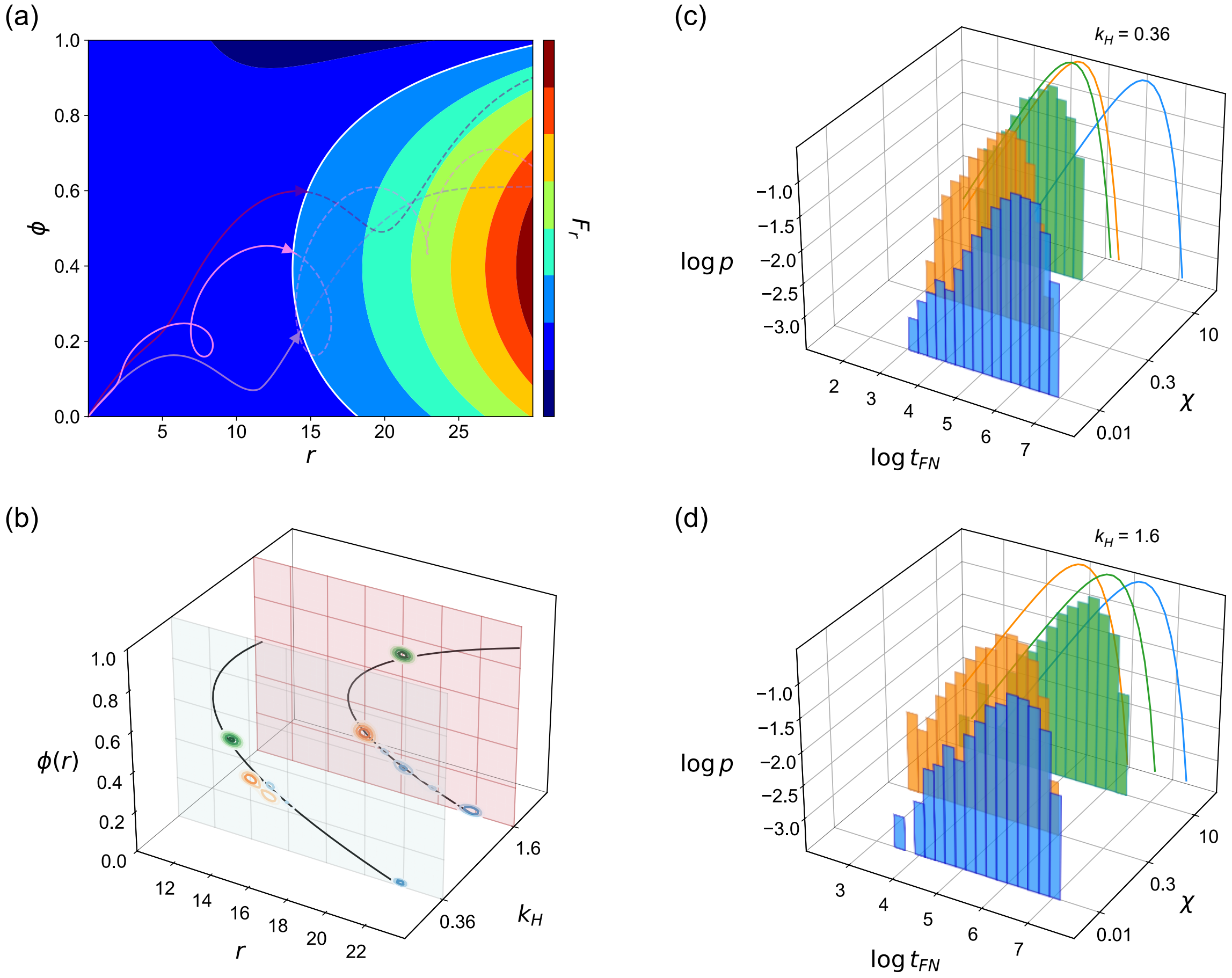

Figure 4. **Overview of the first-passage problem in early-stage nucleation. (a)** Scheme of the first-passage paths within the $(r, \phi)$ phase space. The solid curves represent the first-passage trajectories from the initial state to the critical nucleation regime, superimposed on a background color map indicating the nucleation-driving force $F_r$. **(b)** Joint statistical distributions of the interfacial order parameters $\phi(r)$ and cluster radii $r$ at the onset of the first critical nucleation events with the reordering susceptibility $\chi$ = 0.01 (blue), 0.3 (orange), and 10 (green), respectively. Black lines serve as a guide to the eye. **(c–d)** Distributions of the first-passage times resolved under different conditions. The projected curves on the back planes denote the fitted exponential distributions.

## FIRST-PASSAGE PROBLEM

The methods developed in the study of the first-passage problems are of great importance in non-equilibrium statistical mechanics(*49–52*). Previous theories have mostly used the equilibrium thermodynamics to estimate the nucleation rates and fluxes from the critical free energy(*9–11*). On the contrary, the first-passage methods address the non-equilibrium behaviors. The first-passage paths in early nucleation are the evolution courses of condensed clusters from the beginning to the very first times that they reach the critical nucleation regime (Fig. 4a). This is motivated by two factors: first, the dynamic transition from the early-nucleation to late-growth stages is highly irreversible; second, after the clusters enter the late-growth stage, the dominant mechanism of further condensation is no longer surface adsorption but cluster-cluster coalescence(*18*, *23*, *53*, *54*).

In the stochastic simulations, every complete trajectory covers the full process of a single cluster from its initial condition to the late stable-growth regime. We extracted the first-passage paths from individual evolution trajectories and conducted statistical analysis of the path-dependent properties. Among all the tracked properties, the interfacial order, radius, and time at the ends of first-passage paths are the most important, as they depict the states and times at the first critical nucleation events of targeted clusters. We use the short-form symbols $\phi_{FN}(r)$, $r_{FN}$, and $t_{FN}$ to denote them. Note these quantities do not have direct space-time dependence.

The statistical distributions of the correlated $\phi_{FN}(r)$ and $r_{FN}$ under different conditions are projected onto the $[r, \phi(r)]$ planes (Fig. 4b). As $\chi$ increases from 0.01 to 10, $\phi_{FN}(r)$ increase monotonically. This is a consequence of the enhanced response of the order-parameter field to the reordering-driving force that pushes $\phi$ from the non-reordered limit $\phi_S$ towards the full-reordered states $\phi_{FR}$ which strictly satisfy the condition that $\phi_{FR} > \phi_S$. This situation is different when it comes to $r_{FN}$. With the increase of $\chi$, $r_{FN}$ drop monotonically at weak coupling, but $r_{FN}$ first drop then rise at strong coupling, manifesting the turnover effect caused by the relative shifts of the critical nucleation regime and the pro-nucleation states $\phi_{PN}$ with regards to the full-reordered states when the coupling strength $k_H$ is enhanced (Fig. 2f). When $k_H$ is small, $\phi_{PN}$ are generally higher than $\phi_{FR}$. Thus, the nucleation-driving force changes monotonically as $\chi$ increases from zero to $\infty$. However, when $k_H$ is large, $\phi_{PN}$ are lower than $\phi_{FR}$ as the critical nucleation regime is approached, so $\phi_{PN}$, indicating the turning point of the nucleation-driving force, become intermediate states between the non-reordered limit ($\chi = 0$ and $\phi = \phi_S$) and the full-reordering limit ($\chi = \infty$ and $\phi = \phi_{FR}$).

In a linear-scale histogram, the distribution of $t_{FN}$ can be described by an exponential function; whereas, in a log-scale histogram, a minor shift at the small-$t_{FN}$ region is captured, indicating a transient incubation time within which no critical nucleation event occurs. The distributions under various conditions are presented in Fig. 4c and Fig. 4d. As $\chi$ increases from 0.01 to 10, $t_{FN}$ drop monotonically at weak coupling, but they first drop then rise at strong coupling.

To further quantify these non-equilibrium features, we investigated the path averages of the relative shifts of $\phi_{FN}(r)$ and $r_{FN}$, as well as $t_{FN}$, and applied symbolic definitions as below:

$$\langle \phi^{\circ}_{FN}(r) \rangle_P = \frac{1}{N} \int \mathcal{D}\phi \ \frac{\phi_{FN}(r) - \phi_S}{\phi_{FR|r_{FN}}(r) - \phi_S},$$
$$\langle r_{FN} \rangle_P = \frac{1}{N} \int \mathcal{D}\phi \ r_{FN},$$
$$\langle t_{FN} \rangle_P = \frac{1}{N} \int \mathcal{D}\phi \ t_{FN},$$

where $\int \mathcal{D}\phi$ stand for the summations of all first-passage paths following the spirit of the path integral in field-theoretic methods. The subscripts $P$ on the LHS highlight the path-averaged strategy in contrast to the ensemble-averaged where the summations are taken with all the systems sampled at the same time. Since $\langle t_{FN} \rangle_P$ reflects the expected time cost of critical nucleation, it is used to delineate the nucleation time. The results under various parameter conditions are shown in Fig. 5. When $\chi$ approaches zero, reordering is unlikely to be seen in the observed timescale. In this non-reordered limit, the degrees of freedom of the order-parameter field are reduced and the nucleation process of a single cluster is a 1D heterogeneous random walk along the axis of the radius(*36*). In the opposite limit when $\chi$ approaches infinity, all the observable states are the full-reordered states, hence each evolution trajectory is a minimal-energy path. In these cases, there is no variance or standard deviation in $\phi_{FN}(r)$ and $r_{FN}$ (Fig. 5a and Fig. 5b), but only in $t_{FN}$ (Fig. 5c). When the value of $\chi$ is intermediate, the observed $\phi_{FN}(r)$ denote amorphous states between $\phi_S$ and $\phi_C$. We also computed the path-averaged energy at the first critical nucleation events, denoted by $E_C$ (inset of Fig. 5c). It is clear that $E_C$ cannot characterize $\langle t_{FN} \rangle_P$ as an effective barrier, indicating the breakdown of previous theories.

These path-averaged results from stochastic simulations are compared with a mean-field model where the degrees of freedom associated with the order-parameter field are reduced via a two-step procedure. First, the spatial domain of a multi-shell structure is coarse-grained into a two-component core-shell structure. Second, it is assumed that the effective order parameter on core does not directly contribute to the nucleation dynamics, while the local order parameter on shell is governed by the following equation:

$$\phi(r,t) = [\phi_{FR}(r) - \phi_S]\varphi^\circ + \phi_S, \quad (10)$$

with the dynamic factor

$$\varphi^\circ = 1 - e^{-k_\varphi \chi}, \quad (11)$$

where $k_\varphi$ is a relaxation parameter without space-time dependence. Here $\varphi^\circ$ stands for $\langle \phi_{FN}^\circ(r) \rangle_P$ in a mean-field way. The proposal of this relation is inspired by the reordering dynamics where $F_\phi$ is a linear function of $\phi$. Since $\phi_{FR}(r)$ is determined once $r$ is known and $\phi(r,t)$ is a scalar variable, a short-form notation is applied as $\varphi(r) \equiv \phi(r,t)$ where the time dependence is implicitly absorbed into the radius as $r = r(t)$. Thus, the degrees of freedom of the $\phi$ function are removed, and $r(t)$ is the only relevant time-dependent variable. The mean-field nucleation-driving force is denoted by $F_r^\varphi(r)$, and the critical nucleation regime is reduced to a critical point $r = r_c^\varphi$ on the axis of radius that fulfills $F_r^\varphi(r_c^\varphi) = 0$. The mean-field model gives the solution for the mean first-passage time $t_{FN}^\varphi$ in a simple form:

$$t_{FN}^\varphi = \frac{1}{\mu k_B T} \int_0^{r_c^\varphi} \int_0^r e^{\frac{G(r)-G(r')}{k_B T}} dr'\, dr, \quad (12)$$

with $\mu$ the mobility coefficient, and $G$ the effective nucleation energy:

$$G(r) = -\int_0^r F_r^\varphi(r') dr'. \quad (13)$$

Note that when $\chi \to 0$, $t_{FN}^\varphi$ converges toward a $k_H$-independent constant $t_0$, which corresponds to a non-reordered limit where $\phi = \phi_S$ . The mean-field solutions agree with the simulation results (Fig. 5).

We calculated the effective nucleation energy when the critical nucleation condition is met: $G_C \equiv G(r_c^\varphi)$. It gives a good characterization for $\langle t_{FN} \rangle_P$ in contrast to $E_C$ (Fig. 5c). Therefore, we propose the closed-form approximate solution of $\langle t_{FN} \rangle_P$ as follows:

$$\langle t_{FN} \rangle_P = t_0 \exp\left(\frac{G_C - G_0}{k_B T}\right), \quad (14)$$

where $G_0 = G_{C|\chi=0}$. It is worth emphasizing that $G$ is physically different from $E$, though both are energy functions producing driving forces. Since a single cluster is an open system and adsorption is not governed by a conservative force, $G$ is not a function of state by its physical nature but a path-dependent nucleation energy expressed in a mean-field manner. However, $E$ functions as a potential energy generating a conservative force $F_\phi$ when the system boundary is fixed. As a result, $G_C$ is an effective nucleation barrier remarkably different from those in previous works, as it considers the non-equilibrium properties with the explicit dependence on $\chi$. In this way, this nucleation energy provides a new insight into the evaluation of nucleation barrier.

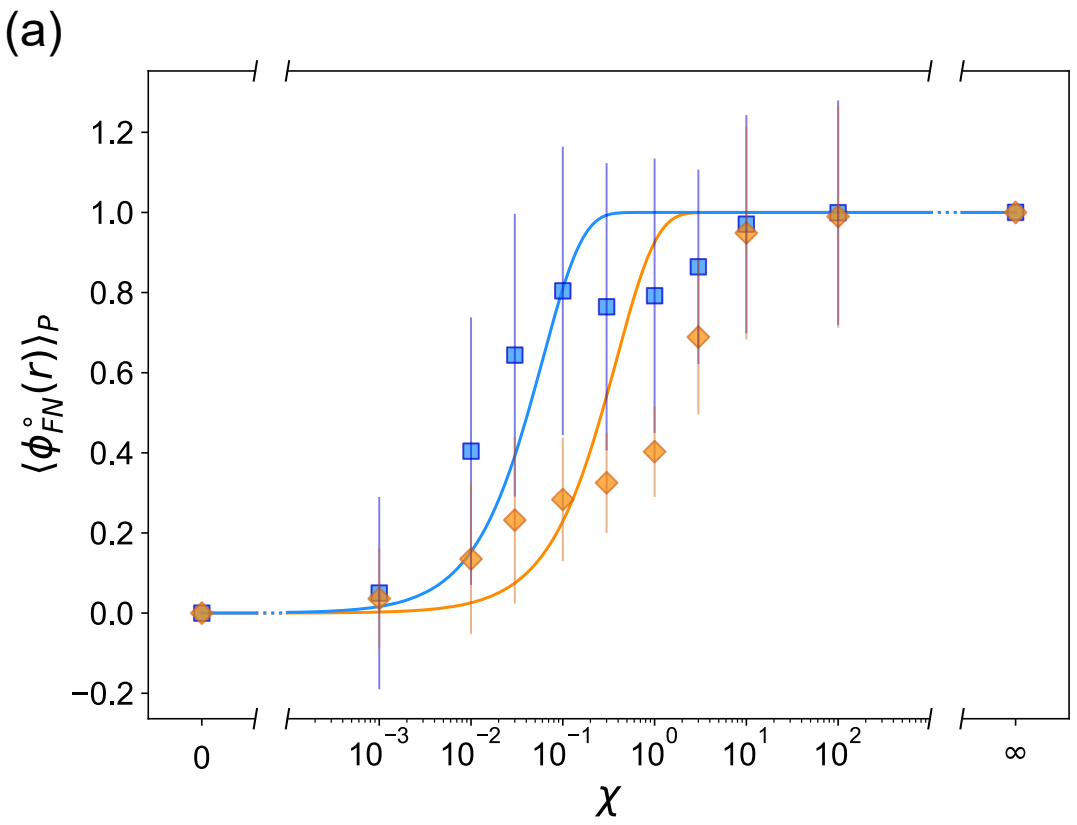

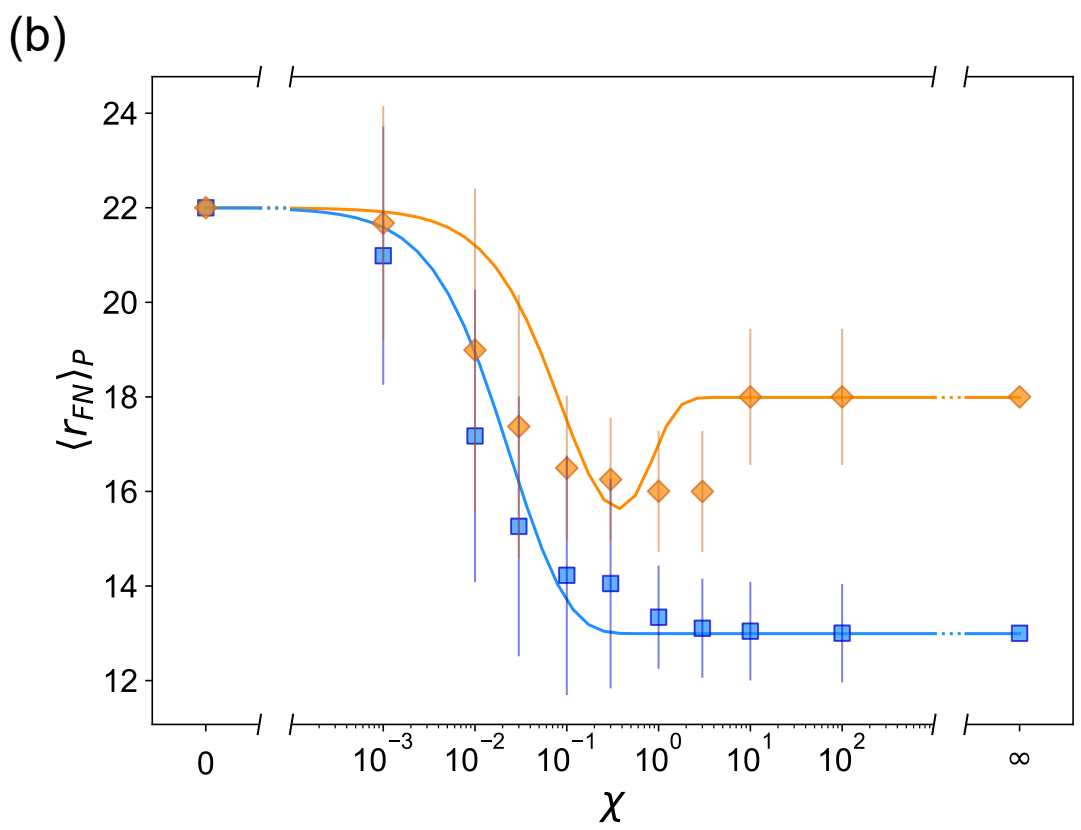

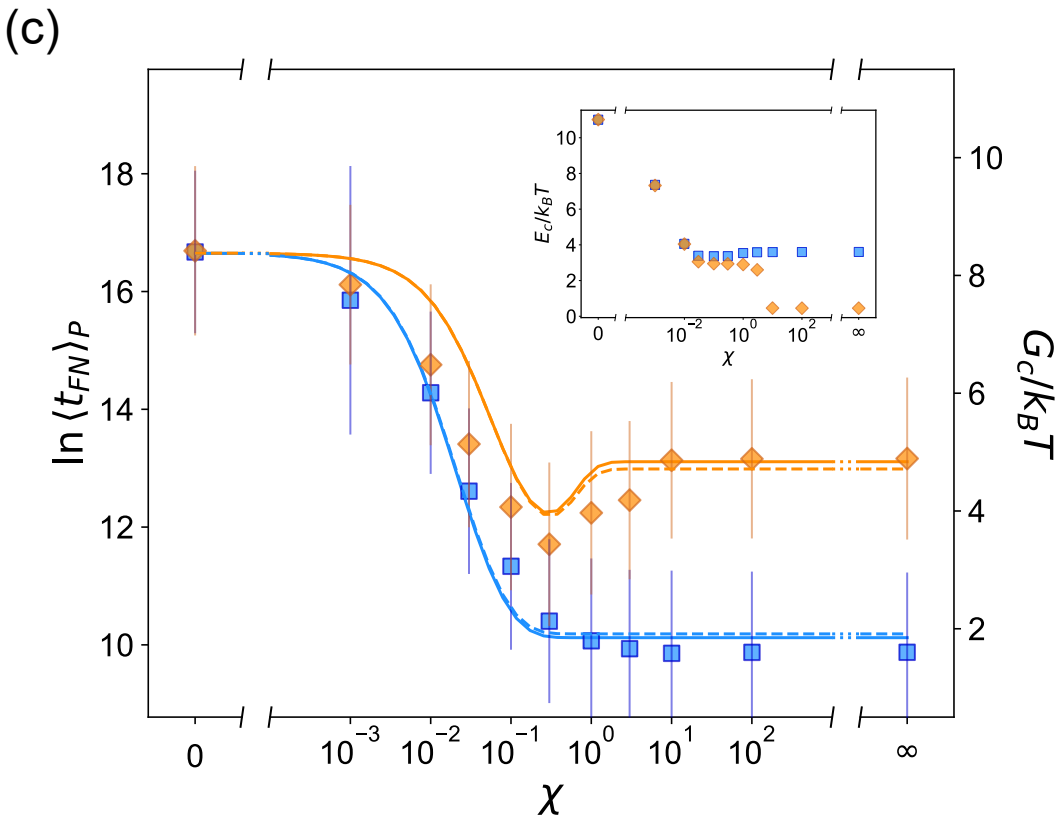

Figure 5. **First-passage properties evaluated across a wide parameter range.** The path-averaged **(a)** relative interfacial order $\langle \phi_{FN}^\circ(r) \rangle_P$, **(b)** cluster radius

$\langle r_{FN} \rangle_P$, and **(c)** transition time $\langle t_{FN} \rangle_P$, with the critical potential energy $E_c$ in the inset, captured at the onset of the first critical nucleation events as $\chi$ varies, with coupling strengths $k_H$ = 0.36 (blue), and 1.6 (Orange). Markers and error bars represent the mean values and standard deviations evaluated from simulations. Curves denote the analytical mean-field solutions. The dashed curves in (c) represent the effective nucleation barrier $G_c$, corresponding to the right vertical axis sharing an identical scale with the left axis.

## DISCUSSION

The multi-shell model developed herein is designed to capture the stochastic reordering-nucleation dynamics of early-stage nucleation. However, its current formulation is not intended for the late-stage growth stage, where cluster-cluster coalescence and diffusion-limited mechanisms become the dominant drivers of domain enlargement(*18*, *23*, *53–55*). Furthermore, as the reaction dynamics for reordering and adsorption are derived in an over-damped limit, the model would require the introduction of a symplectic structure to be applicable to under-damped systems(*56*, *57*). We also assume standard transport mechanism; should anomalous diffusion or active motion prove essential to a specific system(*58–61*), so if the complexity in transportation is essential, the waiting time between adsorption events should be modified accordingly.

The quantitative relation between the nucleation time and the reordering susceptibility highlights the key role of reordering in mediating nucleation kinetics. This connection offers a potential resolution to the discrepancies between contemporary experiments and previous theoretical predictions(*19–21*). Particularly, our model provides a physical principle for the multi-step nucleation pathways observed in nanocrystallization: when the coupling strength $k_H$ is high, pathways characterized by high structural order are effectively prohibited because their associated nucleation times exceed the observable timescale. Consequently, nucleation is restricted to pathways with low structural order, which accurately depicts the multi-step process where intermediate amorphous structures persist throughout the early stage(*15–18*).

The predicted turnover offers a compelling target for future experimental verification. While tuning reordering susceptibility is challenging in nanocrystallization, an equivalent shift can be achieved by manipulating the environment. Since all temporal and rate values are normalized to the adsorption waiting time $\tau_r$, one can effectively shift $\chi$ by controlling concentrations or diffusional properties. In biomolecular systems, direct control of the reordering rate is even more feasible through the modulation of enzymatic activities via chemical or biological approaches(*62–64*).

Finally, while our phenomenological formulation using the order-parameter field $\phi$ is a standard statistical-mechanics treatment in condensed matter physics(*40*, *41*), its application to complex biological systems, which often lack microscopic symmetry, presents a quantification challenge. Nevertheless, rapid advancements in machine-learning techniques provide promising tools(*31*, *65–67*). When the reordering-nucleation process is modeled with atomistic resolution, the machine-learning-based collective variables can serve as a bridge, capturing the microscopic structural changes that underpin our phenomenological description.

## METHODS

**Vectorization**: The state of a condensed cluster is described by $(r, \phi)$ where $r(t)$ is the time-dependent radius and $\phi(x,t)$ is the space-time-dependent order-parameter field. In the multi-shell model, the spatial domain is reduced to discrete layers of shells, so that the order-parameter field is vectorized in the following manner:

$$\phi(x) = [\phi(x_1), \phi(x_2), \dots, \phi(r)],$$

where $x_i$ is the radial coordinate of the outer surface of the $i$-th shell, $r$ is the radius of the cluster, and we omit the temporal coordinate for simplicity. Each shell is set to have the same thickness $l$ which is the unit length, hence $x_i = i$ and $r = i_{max}$. By this means, the gradient is expressed in a discrete fashion as follows:

$$\nabla\phi(x_i) = \phi(x_{i+1}) - \phi(x_i),$$

for $x_i < r$, and

$$\nabla\phi(x_i) = \phi(x_i) - \phi(x_{i-1}),$$

for $x_i = r$ . Different from the continuous-space formalism, $\delta/\delta\phi|_{x=x_i} = \partial/\partial\phi(x_i)$ in the multi-shell model.

Since the reordering dynamics is modeled as a continuous-time process, the related shift of the order-parameter field is denoted by $d\phi$ , whose vector representation is expressed as below:

$$d\phi = [d\phi(x_1), d\phi(x_2), \dots, d\phi(r)].$$

On the other hand, the change of $\phi$ induced by adsorption is realized by adding an extra component or removing the last component, in a fashion like the direct sum of vectors in algebra. This is symbolically written as follows:

$$\phi \oplus \Delta\phi = [\phi(x_1), \phi(x_2), \dots, \phi(r), \phi(r+l)],$$

in a forward adsorption event where $r \to r + l$, and

$$\phi \ominus \Delta\phi = [\phi(x_1), \phi(x_2), \dots, \phi(r-l)],$$

in a backward adsorption event where $r \to r - l$. Based on the specific adsorption mechanism studied in this paper, $\phi(r+l)$ is always equal to $\phi_S$ in a forward adsorption step.

**Stochastic simulation**: With the vector representation of the state, the reordering-nucleation dynamics of a single condensed cluster was simulated in finite-difference methods. The elementary timestep was set to $\tau_\phi$, and the state was updated per step based on reordering. The raw values of $\tau_r$ were set as multiples of $\tau_\phi$, and whether and how the state was updated by adsorption after every $\tau_r$ were determined through a stochastic method. The final timescales were normalized in relation to $\tau_r$.

In this study, we inspected the different coupling conditions when $k_H$ equals 0.36 (weak) and 1.6 (strong). For each coupling case, eleven dynamic conditions were analyzed by tuning $\chi$, $\tau_\phi$, and $\tau_r$, while other parameters were kept constant. With each set of parameters, we ran 1000 parallel simulations for the individual clusters. The initial conditions were set as $r(t=0) = 1$, and $\phi(x=1, t=0) = \phi_S$. Every simulation was terminated when $r$ reached 50. After the simulations finished, we examined the last times and ensured that all clusters entered the post-critical regime where the nucleation-driving force was positive.

## ACKNOWLEDGMENTS

The authors acknowledge support from the HKUST Central High Performance Computing Cluster. This work was funded by RGC grants C2001-23Y and C6004-22Y.

## COMPETING INTERESTS

The authors declare no competing interests.